\documentclass[]{spie} 

\usepackage{amsmath}
\usepackage{graphicx}
\usepackage{amsfonts}
\usepackage{amssymb}%
\providecommand{\U}[1]{\protect\rule{.1in}{.1in}}

\title{On the relationship between speech and hearing}

\author[a]{Srinivasan Umesh}
\author[b]{Leon Cohen}
\author[c] {Douglas Nelson}
\affil[a]{Dept. of Electrical Eng.,IIT Madras-600042, India}
\affil[b]{City University of New York, 695 Park Avenue, New York, NY 10021}
\affil[c]{U.S. Dept. of Defense (retired), Ft. Meade, MD 20755}

\authorinfo{Author Umesh Srinivasan was supported in part by the Department of
Science \& Technology, Ministry of Science \& Technology, India under project
SR/S3/EECE/0008/2006-SERC-Engg.}

\begin{document}
\maketitle
\begin{abstract}
 We present a framework for experimentally linking speech production and
hearing. Using this approach, we describe experimental results, that lead to
the concept that sounds made by different individuals and perceived to be the
same can be transformed into each other by a \textquotedblleft speech
scale\textquotedblright. The speech scale is empirically determined using
\emph{only }speech data. We show the similarity of the speech scale to the MEL
scale of Stevens and Volkmann, which was derived \emph{only} from hearing
experiments. We thus experimentally link speech production and hearing.   
\end{abstract}

\section{Introduction}

For speech communication, nature has provided both a speech \textquotedblleft
transmitter\textquotedblright\ and \textquotedblleft
receiver\textquotedblright; therefore, speech production and hearing must be
highly coupled systems. While a particular individual has only one receiver
(the ear), one has to understand and classify sounds from many different
transmitters (speakers). This is because, when different individuals enunciate
sounds which are perceived to be the same, the actual acoustical waveforms are
different, the corresponding spectra are different, and the location and
separation of the formants are different. However, since they are perceived to
be the same, the enunciations must have a commonalty that the ear extracts in
recognizing the similarity of the sounds. We call this the speech-hearing
connection and the understanding of this issue the speech-hearing problem.

In 1940, Stevens and Volkmann established the fundamental psychoacoustic
response of the ear, known as the MEL scale \cite{Stevens-40a}. It relates
perceived frequency, $f_{mel}$, and the actual physical frequency, $f$, of the
sound wave. The MEL scale is solely based on psychoacoustic \textit{hearing}
experiments. We have performed\emph{ speech} experiments aimed at showing the
common attributes of speech signals of perceptually similar sounds. Our
experiments establish a \textquotedblleft speech scale\textquotedblright,
which is based \emph{solely} on speech production and has led us to the
following view of the speech-hearing connection. The spectra of sounds made by
different individuals and perceived to be phonologically identical can be
transformed into each other by a universal frequency-warping function. The
transformation results in identical spectra, except for a speaker dependent
translation factor. We call the universal frequency-warping function the
\emph{speech-scale}. We show that the experimentally obtained
\textquotedblleft speech scale\textquotedblright\ is similar to the MEL scale
and, therefore, we argue that we have experimentally linked speech and
hearing. We emphasize that our speech scale was obtained solely from speech
experiments and in no way used any results of the psychoacoustic MEL scale or
any other hearing data.

The organization of the paper is as follows. In the next section we give an
simplified overview of our approach and the basic concepts we have used to
establish a "speech scale". In section 3 we describe the mathematical issues
and the details of the method used to estimate speech-scale. In Section 4 we
discuss the similarity between speech-scale and mel-scale. Discussion and
conclusions are made in Section 5.

\section{An overview of the approach}

\begin{figure}[ptb]
\begin{center}
\includegraphics[width=4.0in]{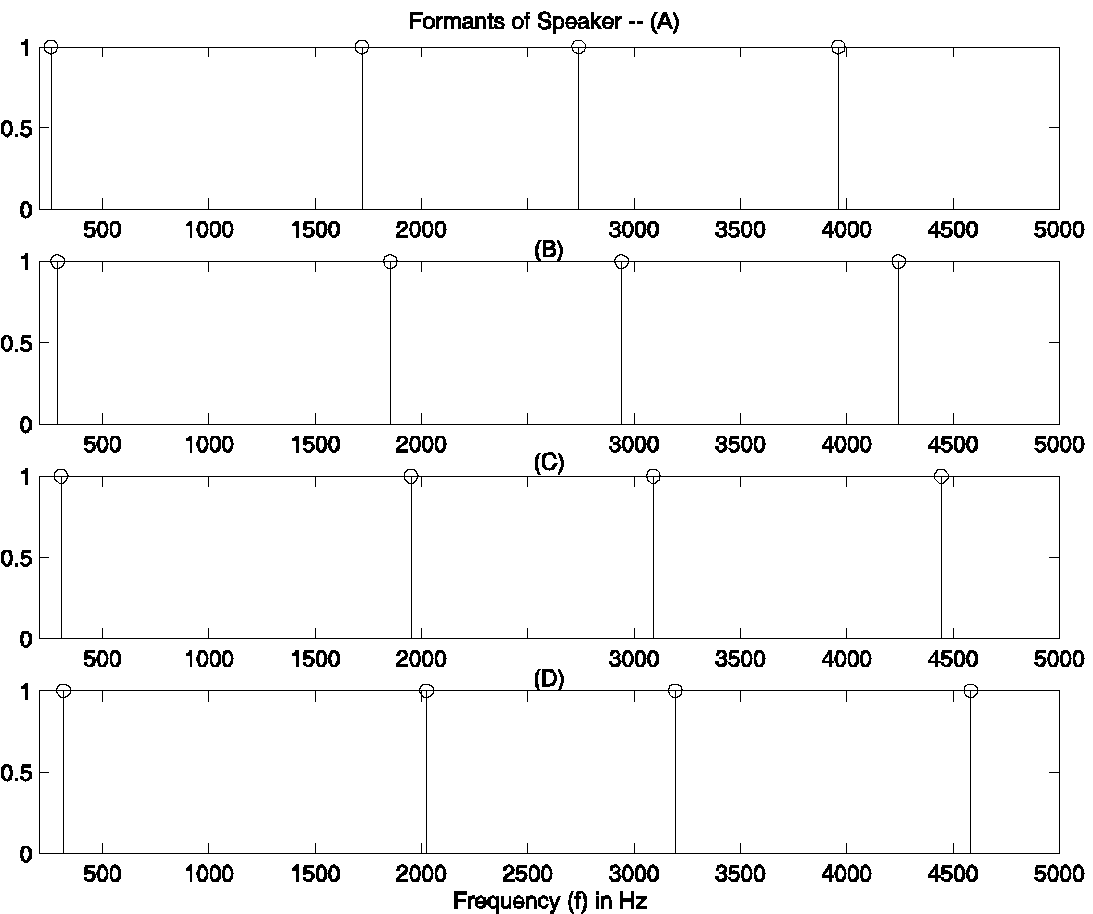}
\end{center}
\caption{The figure shows the formants from four different hypothetical
speakers. The horizontal axis is real frequency, $f$ measured in Hertz.}%
\label{norm_aa}%
\end{figure}

We first explain the basis and motivation of our approach with a simple
idealized example. Suppose, hypothetically, that we have the utterances of
four persons enunciating a particular vowel, say /aa/. We may say that the
vowel is the same in each instance if critical listeners agree that they hear
the same sound for each of the utterances.
Now suppose we take the spectra, that is Fourier transform, of each person's
utterances and plot them as in Fig. \ref{norm_aa}, where for simplicity, we
only plot the formants (resonant frequencies).
For the case of illustration here we have generated these formants for /aa/ by
using the two-tube model for speech production \cite{Flanagan-72a}.
\footnote{In a two-tube model $L_{1},A_{1}$ represents the length and area of
the \textquotedblleft pharyngeal\textquotedblright\ cavity or the part
starting from the vocal cords at the back of the mouth and $L_{2},A_{2}$
represents the length and area of the \textquotedblleft oral\textquotedblright%
\ cavity, the part ending at the lips or the front part of mouth. The ratio of
the pharyngeal cavity to the oral-cavity lengths, that is $\frac{L_{1}}{L_{2}%
},$ is largest for men, intermediate for women and smallest for children.
Usually only ${L_{2}}$ changes significantly from speaker to speaker, while
${L_{1}}$ is similar among different speakers. Therefore, in our example, we
have maintained $L_{1}$ and the ratio of areas constant, while changing the
lengths of $L_{2}$. The lengths of ${L_{2}}$ have been appropriately varied so
that the corresponding formants correspond to that of different speakers.}

\begin{figure}[ptb]
\begin{center}
\includegraphics[width=4.0in]{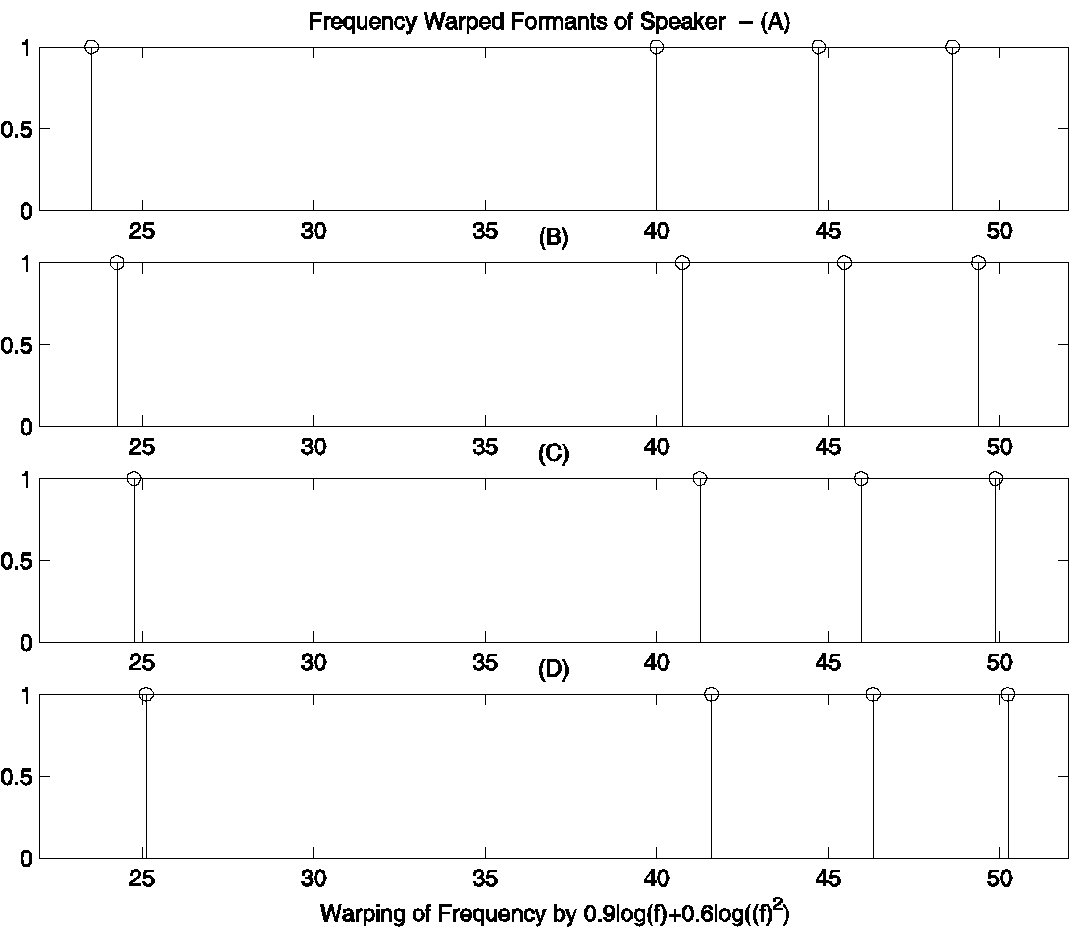}
\end{center}
\caption{Each of the spectra in Fig. 1 is transformed according to the
following function $\nu=0.9\log(f)+0.6(\log(f))^{2}$ and plotted respectively.
The horizontal axis now is $\nu$. In the new domain, $\nu$, the spectra are
identical except for a translation factor. See next figure.}%
\label{norm_aa_nu_1}%
\end{figure}

Now, as can be seen in Fig. \ref{norm_aa}, the spectra, that is, the formant
frequency locations are different, but, since the utterances are perceived to
be the same, they must have something in common. What is this communality? In
other words, what do these spectra have in common? We emphasize that the
spectra are not merely translated versions of each other as that would make
the issue trivial. Indeed if they were translated versions of each other that
could be readily determined in a variety of ways. If they are not translated
versions of each other, then one may ask whether there exists a transformation
of each spectra that would indeed make them identical or possibly translated
versions of each other. Such a transformation would be non-linear. Such
transformations are often called warping functions because one can think of a
non-linear transformation as a warping of the axis. What is important here is
to emphasize that, if there exists such a transformation, then it must be
universal to be of any significance. By universal we mean that the \emph{same}
transformation is used for each speaker and it must not be adjusted for
different speakers. For illustration purposes we now do precisely that for the
above formant example and take each spectrum and warp the frequency-axis by a
function which we call the \textquotedblleft universal warping
function\textquotedblright. We denote the frequency-warping function as
$\nu=g(f)$, and for this illustrative example we have used the function
$\nu=0.9\log(f)+0.6(\log(f))^{2}$. The result of applying this function to
each of the spectra is illustrated in Fig. \ref{norm_aa_nu_1}. We again note
that we use the term \textquotedblleft universal warping
function\textquotedblright\ to emphasize the fact that the \emph{spectra} are
transformed by the \emph{same} warping function irrespective of the speaker.
In the next section, we will discuss how one arrives at such warping functions
for idealized cases and more importantly in Section~4 we will describe how one
can empirically obtain such a function for real speech. We also emphasize here
that whether such a universal function exists or not is an empirical question
and can only be ascertained by the real data.

\begin{figure}[ptb]
\begin{center}
\includegraphics[width=4.0in]{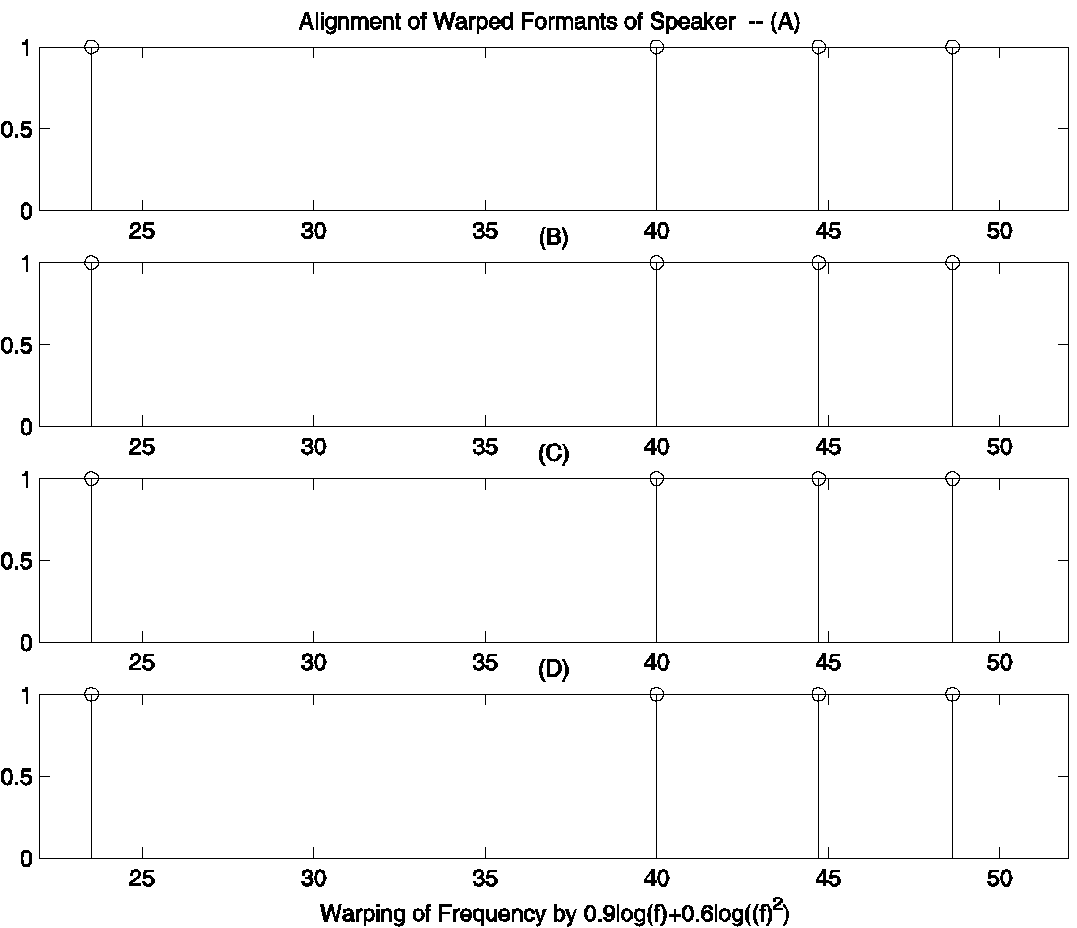}
\end{center}
\caption{This plot confirms that after aligning the formants they are indeed
identical.}%
\label{norm_aa_nu_2}%
\end{figure}

Now examine Fig.~\ref{norm_aa_nu_1}. It can be seen that:

a) the transformed spectra are merely shifted versions of one and another and
hence we have exposed the commonalty of the original spectra of Fig.
\ref{norm_aa}. To emphasize this, we align the spectra with the first line, as
illustrated in Fig.~\ref{norm_aa_nu_2}. Now one clearly sees that indeed they
are identical. (We note that in general there are many ways to determine
whether spectra are shifted versions of each other, the most fundamental way
is to note that the absolute value of the Fourier transform of a function and
a translated version of a function are equal.)

b) the shift is different for each transformed spectrum, and, hence, the shift
is speaker dependent.

We again emphasize that although Fig.\ref{norm_aa} looks similar to
Fig.\ref{norm_aa_nu_1}, it can be easily verified that simple shifting (or
scaling) will not align the spectra of Fig.\ref{norm_aa}. It is only after the
application of the \textquotedblleft universal warping
function\textquotedblright\ that the spectra appear as shifted versions of
each other. Of course, for real speech, we do not know whether such a
\textquotedblleft universal-warping function\textquotedblright\ exists, and,
if it does exists, we do not know the form of the universal warping function.
The purpose of this article is to describe experimental evidence indicating
that, indeed, such a universal warping function exists and to develop the
numerical procedures to obtain it from real speech.

\section{Mathematical and Experimental issues}

Our basic idea is that the spectra of sounds made by different individuals and
perceived to be the same can be transformed into each other by a universal
warping function and that the transformation results in identical warped
spectra except for a translation factor. We denote the transformation from
physical frequency, $f$, to an alternate domain $\nu,$ as the
\emph{speech-scale}, and write%
\begin{align}
\nu &  =g(f)\hspace{1in}(\text{speech scale})\\
f  &  =h(\nu)\hspace{1in}(\text{inverse speech scale})
\end{align}

Suppose we have two enunciations by different speakers, $A$ and $B$, with
corresponding spectra $P_{A}(f)$ and $P_{B}(f)$. We warp each spectrum with
the universal function, $\nu=g(f)$, to obtain two new spectra $S_{A}(\nu)$ and
$S_{B}(\nu)$ defined by the following procedure:
\begin{align}
S_{A}(\nu)  &  =P_{A}(h(\nu))\\
S_{B}(\nu)  &  =P_{B}(h(\nu))
\end{align}
then we claim that, for enunciations which are perceived to be the same, the
warped spectra are related by%
\begin{equation}
S_{A}(\nu)=S_{B}(\nu+\alpha_{AB})
\end{equation}
where $\alpha_{AB}$ is a constant. In general, we use the notation that
spectra in physical frequency are represented as $P(f)$ and spectra in the
warped domain are represented as $S(\nu)$. If the claim is correct, the
question arises how can one show it and how can one determine the universal
transformation that we call the speech scale. Before describing our approach,
we must emphasize two things. First, even though we have used two speakers in
the above formulation, it is crucial to understand that this must be true for
\emph{all} speakers. That is, it is crucial that the \emph{same}
universal-warping function $\nu=g(f)$ is applicable to all speakers and all
phonemes or speech sounds. Secondly, whether this view and mathematical
relationship are true, or not, can only be determined experimentally using
actual speech data.

We have been motivated by Fant's seminal work \cite{Fant-59a,Fant-75a} where
he considers relationship between formants of different speakers. First, we
discuss a very simplified model where we model the vocal tract as a uniform
tube. For a uniform tube the spectra (or formant frequencies) are inversely
proportional to the length of the tube. Hence for different speakers having
different tube-lengths, the spectra are related by
\begin{equation}
P_{A}(f)=P_{B}(\kappa_{AB}f) \label{eq:linmod}%
\end{equation}
where $\kappa_{AB}$ is the ratio of the lengths of the tubes for the two
speakers. Recalling that multiplication becomes addition after the application
of log operation, the universal warping function for this simplified model is
$\lambda=\log(f)$. This can be easily verified since
\begin{equation}
S_{A}(\lambda)=P_{A}(f=e^{\lambda})=P_{B}(\kappa_{AB}e^{\lambda}%
)=P_{B}(e^{\lambda+\ln\kappa_{AB}})=S_{B}(\lambda+\ln\kappa_{AB})
\label{eq:log_shift}%
\end{equation}

Fant \cite{Fant-59a,Fant-75a} pointed out that Eq.~\ref{eq:linmod} is a very
crude model and has observed that the scaling factor $\kappa_{AB}$ changes for
different formant frequencies for any two speakers. Motivated by Fant's
observation we model the relationship between spectra of two speakers by the
following piece-wise linear model
\begin{equation}
P_{A}(f)=P_{B}(\gamma_{AB}^{\beta_{l}}f)\;\;\;\;\mathrm{where}\;f\;\in
\;[L_{l},U_{l}] \label{eq:piece}%
\end{equation}
where $L_{l},U_{l}$ are lower and higher frequencies of a frequency region.
The parameter $\gamma_{AB}$ depends only on the pair of speakers and not on
frequency. $\beta_{l}$ is a constant that is \emph{independent} of speakers
and depends on only the frequency region $[L_{l},U_{l}]$. We have estimated
the frequency-dependent parameter $\beta_{l}$ by doing least-squares fit on
\emph{actual} speech data. In all of the experiments in this paper, we have
used the Hillenbrand speech (vowel) database\cite{hil}. The piece-wise linear
approximation to universal-warping function $\nu$, is then given by
\begin{equation}
\nu=\frac{1}{\beta_{l}}\ln(f)\;\mathrm{for}\;f\;\in\;[L_{l},U_{l}]
\end{equation}
The use of this universal-warping function $\nu$ helps separate the
speaker-dependent term as a translation factor and show the commonality in the
speech spectra from different speakers. We refer to this universal-warping
function as the \emph{speech-scale}

\section{Relationship between MEL scale and the speech scale.}

\begin{figure}[ptb]
\begin{center}
\includegraphics[trim={0.5cm 0 0 0.5cm},clip,width=5in,height=3in]{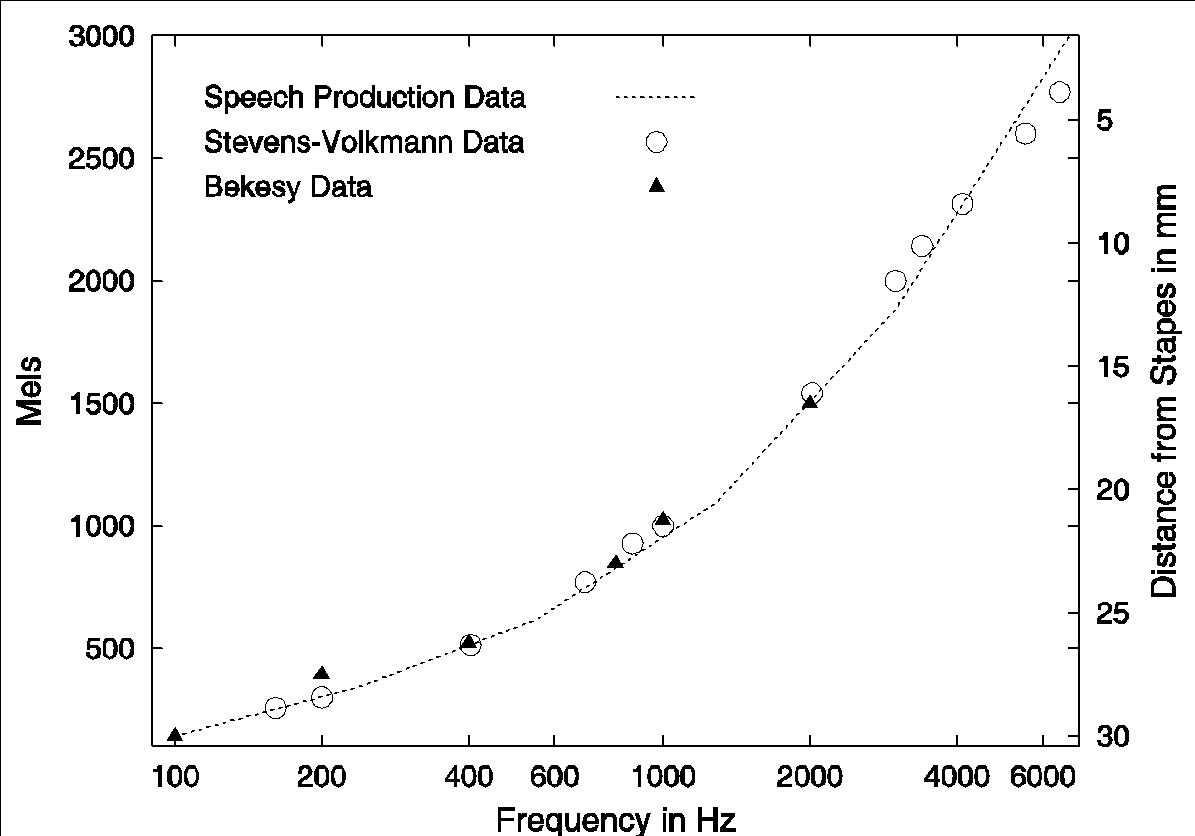}
\vspace{-0.1cm} \vspace{-0.25cm}
\end{center}
\caption{The figure shows the Speech-Scale, Stevens and Volkmann data and
B\'{e}k\'{e}sy's data. The Speech-scale has been obtained empirically from
actual speech data. The Steven and Volkmann data have been obtained from
psycho-physiological study, while the B\'{e}k\'{e}sy data has been obtained
from experiments on the basilar membrane. The fact that all the three curves
are similar shows a strong connection between speech and hearing.}%
\label{consolidated-curve}%
\end{figure}

Stevens and Volkmann~\cite{Stevens-40a} experimentally obtained a non-linear
mapping between perceived and physical frequency of a tone and referred to it
as the mel-scale. We have obtained a speech scale which has been purely
estimated only from speech data with the primary purpose to show the
commonality $between$ speakers for the same utterance. In
Fig.~\ref{consolidated-curve} we plot both the speech scale and Stevens and
Volkmann~\cite{Stevens-40a} data, that is, the MEL scale. (In addition we have
also plotted the von B\'{e}k\'{e}sy basilar membrane data but this issue will
be discussed in the conclusion.) From the figure, it can be seen that the two
curves are remarkably similar. We reiterate, that one has obtained from
psychoacoustic experiments and the other only from speech data. The fact that
the two curves are remarkably similar experimental shows the connection
between speech and hearing.

\begin{figure}[ptb]
\par
\begin{center}
{\includegraphics[trim={0.1cm 0 0 0}, clip,width=6in,height=4.5in]{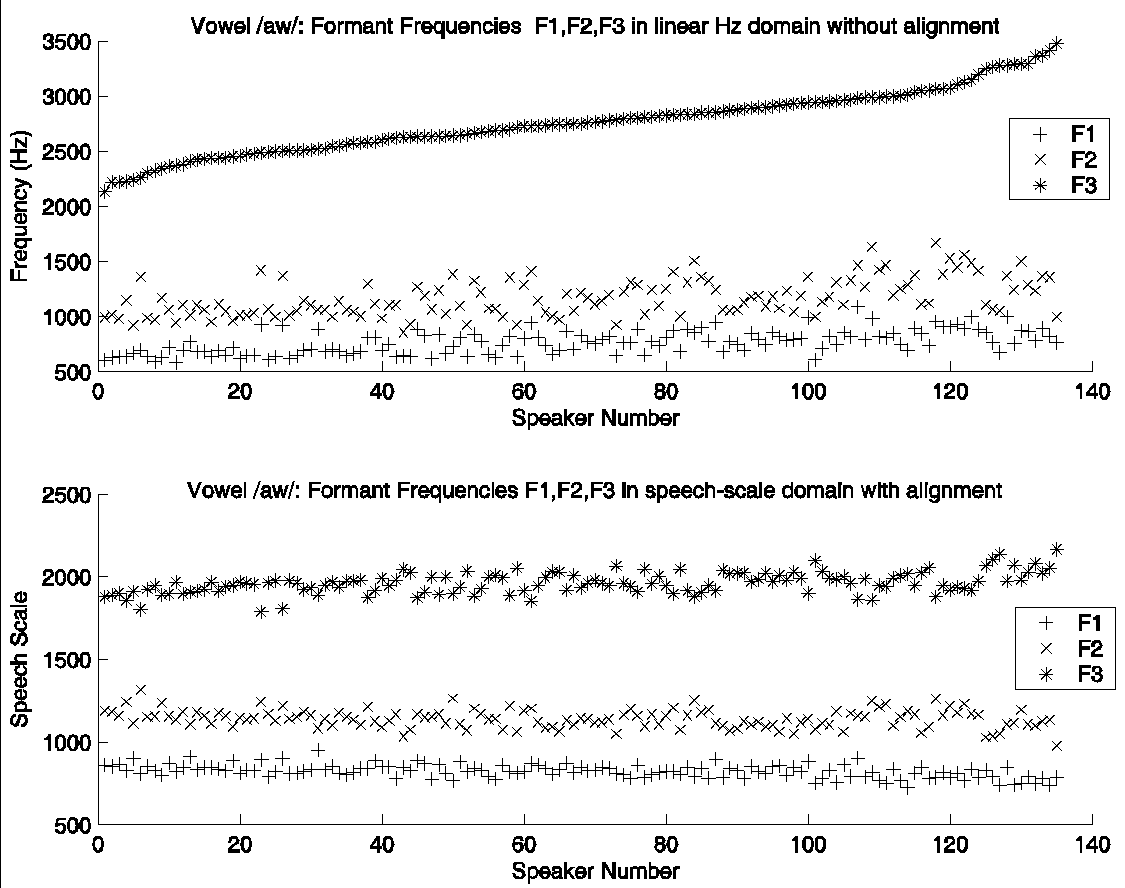}}
\vspace{-0.1cm} \vspace{-0.25cm}
\end{center}
\par
\label{Hil_AW}\end{figure}

To further show the similarities between the the speech scale and MEL\ scale,
we point out that the widely accepted closed-form approximations to mel-scale
have the functional form
\begin{equation}
\eta=a\log_{10}\left(  1+\frac{f}{b}\right)  \label{Eq:Mel-Func}%
\end{equation}
where $f$ is in Hz and $\eta$ is in \emph{Mels}. Here we have used the
mel-formula which is most commonly used in speech and defined with $a=2595$
and $b=700$, i.e.
\begin{equation}
\eta_{\text{MEL}}=2595\log_{10}\left(  1+\frac{f}{700}\right)
\end{equation}
We now take our speech scale and fit it to Eq.~(\ref{Eq:Mel-Func} (with
fitting accuracy of $99.5\%$) to obtain continuous warping function
$\Omega(f)$. The warping function is given by
\label{Eq:Pwwarpgrp}%
\begin{equation}
\eta_{\text{speech}}=2478.24\log\left(  1+\frac{f}{641.94}\right)  .
\label{Eq:Pw_hil}%
\end{equation}
Thus we see that indeed $\eta_{\text{speech}}$ is approximately equal to
$\eta_{\text{MEL}}.$

As further evidence of the ability of the speech scale to extract the
commonality among different speakers enunciating the same sound, we apply it
to study the formants of different speakers for the vowel /AW/. In
Fig.~\ref{Hil_AW}a, we show the natural frequencies of three formants of /AW/
for each of the 139 different speakers in the Hillenbrand database. For each
speaker, the three formant frequencies have been marked by different symbols
along the $y$ axis. As seen, there is considerable variability in the formant
frequencies among different speakers. We now apply the speech scale to the
data of Fig.~\ref{Hil_AW}a and translate each of the resulting spectra to
obtain the best overall alignment \cite{Nelson_SPIE}. The results of this
process are shown Fig.~\ref{Hil_AW}b. As seen, after the application of the
\emph{speech scale}, the formants are aligned remarkably well showing the
commonality among speakers for the vowel /AW/. In a future paper we will
present results for all the vowels.

\section{Discussion and Conclusion}

As we have noted, speech and hearing must be a highly coupled. If one could
demonstrate a relationship between the results of hearing experiments and the
results of speech experiments, a fundamental experimental link would be
established. We believe that we have done that. Using real speech data and
only speech we have experimentally obtained a \textquotedblleft speech
scale\textquotedblright, which is a warping of the speech spectrum that
exposes the communality of sounds which are perceived to be the same. Our
experiments are analogous to the basic psychoacoustic scale obtained by
Stevens and Volkmann \cite{Stevens-40a} which was obtained solely on hearing
data. The fact that they are almost the same and derived totally
independently, one from speech and the other from hearing shows the coupling
or communality between speech and hearing and experimentally establishes the
speech-hearing connection. In addition there is another scale that is of
relevance which we now discuss. It is the frequency to physical place mapping
on the basilar membrane which is based entirely on physical aspects of the
basilar membrane. This
place map, representing the behavior of the basilar membrane as a function of
frequency was discovered by von B\'{e}k\'{e}sy \cite{Bekesy-51a} in his
studies, for which he won a Nobel prize. In his experiments the stapes was
vibrated with a constant amplitude sinusoid, and the frequency response at
various points along the basilar membrane was examined under the microscope.
The mapping relating the frequency of the stimulation to the position of
maximum response on the basilar membrane was thus established. Is there a
relationship between the Mel scale and the
place map of von B\'{e}k\'{e}sy?
This question was addressed by von B\'{e}k\'{e}sy, himself, who noticed the
fundamental significance of the fact that they are similar and noted
\textquotedblleft If we plot the distance from the stapes of the point of
maximum displacement versus frequency, the curve obtained has the general form
of the curves inferred from psycho-physiological data\textquotedblright.

We now compare the speech scale, the Mel scale and the place map of von
B\'{e}k\'{e}sy. The results are shown in Fig.~\ref{consolidated-curve}. The
fact that the three, independent, experimentally derived scales, namely the
place map, the MEL scale and the speech scale, are so similar is an indication
that they may be the basic fundamental experimental linking of the
speech-hearing connection.



\end{document}